\documentclass[a4paper,11pt]{article}
\usepackage{a4wide}
\usepackage{amsmath}
\usepackage{amssymb}
\usepackage{float}
\usepackage[dvips]{graphics}
\usepackage[pdftex]{graphicx}

\newcommand{\be}{\begin{eqnarray}}
\newcommand{\ee}{\end{eqnarray}}

\input{tcilatex}

\begin{document}

\title{The Value of the Cosmological Constant }
\author{John D. Barrow$^{1}$ and Douglas J. Shaw$^{2}$ \\
DAMTP, Centre for Mathematical Sciences, \\
Cambridge University, Cambridge CB3 0WA, \\
United Kingdom\\
}
\maketitle

\begin{abstract}
We make the cosmological constant, $\Lambda $, into a field and restrict the
variations of the action with respect to it by causality. This creates an
additional Einstein constraint equation. It restricts the solutions of the
standard Einstein equations and is the requirement that the cosmological
wave function possess a classical limit. When applied to the Friedmann
metric it requires that the cosmological constant measured today, $t_{U}$,
be $\Lambda \sim t_{U}^{-2}\sim 10^{-122}$, as observed. This is the
classical value of $\Lambda $ that dominates the wave function of the
universe. Our new field equation determines $\Lambda $ in terms of other
astronomically measurable quantities. Specifically, it predicts that the
spatial curvature parameter of the universe is $\Omega _{\mathrm{k0}}\equiv
-k/a_{0}^{2}H^{2}=-0.0055$, which will be tested by Planck Satellite data.
Our theory also creates a new picture of self-consistent quantum
cosmological history.
\end{abstract}

The cosmological constant, $\Lambda $, has played a stimulating role in
gravitation theory ever since Einstein introduced it in 1917 to provide the
gravitational repulsion needed to support a static universe. A new insight
emerged in 1934 when Lema\^{\i}tre \cite{lem} first showed how to
reinterpret it as a Lorentz-invariant vacuum `fluid' in Einstein's
equations. More recently, this formulation has led to its interpretation as
the vacuum energy\ density of the universe \cite{units}, $\rho
_{vac}=\Lambda /8\pi $, as Lema\^{\i}tre suggested, \cite{zeld, wein}.
Before 1998, there was no direct astronomical evidence for $\Lambda $ and
the observational upper bound was so strong\ -- $\Lambda <10^{-120}$ Planck
units -- that many particle physicists suspected that some fundamental
principle must force its value to be precisely zero.

Alas, no such principle was forthcoming. Worse still, any attempt to set $%
\Lambda $ to zero at the start of the universe was overcome by the
generation of large effective $\Lambda $ values when the universe cooled
through phase transition in its early stages. The result was a $\Lambda \ $%
value today at least $10^{56}$ times larger than observations permitted.
Then, in 1998, two independent groups, led by Riess and Perlmutter \cite{obs}
used Type 1a supernovae to show that the universe is accelerating. This
discovery provided the first direct evidence that $\Lambda $ is non-zero,
with $\Lambda \sim $ $1.7\times 10^{-121}$ Planck units.

This remarkable discovery highlighted the question of \textit{why} $\Lambda $
has this unusually small value. It is $10^{121\text{ }}$times larger than
the `natural' value for the vacuum energy of the universe. Moreover, it is
very close to the largest value ($\sim 10^{-120}$) that it could take
without preventing galaxies from having formed \cite{btip}. So far, no
explanations have been offered for the proximity of the $\Lambda $ to $%
1/t_{U}^{2}\sim 1.6\times 10^{-122}$, where $t_{U}$ $\sim 8\times 10^{60%
\text{ }}$is the present expansion age of the universe in Planck time units.
Attempts to explain the coincidence that $\Lambda \sim 1/t_{U}^{2}$ have
relied upon ensembles of possible universes, in which all possible values of 
$\Lambda $ are found. Anthropic selection is combined with some the prior
probability distribution for $\Lambda $ over the ensemble to find the most
probable value that allows galaxies to form \cite{ef}. Clearly, it would be
much more attractive to predict $\Lambda $ directly using a testable theory
without appeal to a multiverse of possibilities. This we shall now do.

We will extend the Einstein-Hilbert variational principle for general
relativity (GR) by promoting $\Lambda $ from being a parameter to a `field'
and only include causal variations which are on and inside our past light
cone. In addition to the usual Einstein equations, this creates a new
integral field equation which determines $\Lambda $ in terms of other
properties of the observed universe. Crucially, the observed classical
history always has\ $\Lambda \sim 1/t_{U}^{2}$ when observed at time $t_{U}.$
In GR, $\Lambda $ is a true constant and is \emph{not} seen to evolve,
although the constant value it takes depends on the cosmic time of
observation. Hence, the resulting history is indistinguishable from GR with
a constant value of $\Lambda $ put in by hand. Moreover, this theory
produces a firm prediction for $\Lambda $ in terms of other measurable
quantities and is testable by future observations.

Conventionally, $S_{\mathrm{tot}}[g_{\mu \nu },\Psi ^{i},\Lambda ;\mathcal{M}%
]$ is the total action of the universe defined on a spacetime $\mathcal{M}$,
with boundary $\partial \mathcal{M}$. $S_{\mathrm{tot}}$ is a functional of
the metric, $g_{\mu \nu }$, and matter fields, $\Psi ^{i}$, and also depends
on the fixed parameters like $\Lambda $ and the fundamental constants. It is
well known that the classical field equations follow from extremizing $S_{%
\mathrm{tot}}$. At the quantum level, physics is determined by a partition
function $Z_{\Lambda }[\mathcal{M}]$ which is a sum of $\exp (iS_{\mathrm{tot%
}})$ over all configurations of the fields with parameters such as $\Lambda $
fixed: 
\begin{equation*}
Z_{\Lambda }[\mathcal{M}]=\sum_{g_{\mu \nu },\Psi ^{i}}e^{iS_{\mathrm{tot}}}.
\end{equation*}%
The dominant configurations, or histories, are those in which the classical
field equations hold.

Our new proposal for solving the $\Lambda $ problem requires only a simple
modification of this standard variational principle. When we promote $%
\Lambda $ from a fixed parameter to a field that can take many possible
values, all of which contribute to the partition function, we also demand
that the action and the partition function are causal, so they only depend
only field configurations in the observer's causal past. This preserves
classical causality.

This promotion of $\Lambda $ from fixed parameter to one that can take many
values occurs in some theories of \ quantum gravity. In string theory, $%
\Lambda $ can take many possible values over a continuous, or tightly-spaced
discrete spectrum. There may be $O(10^{500})$ different vacua and $\Lambda $
values with a spacings of $O(10^{-500})$ in Planck units. Provided this
spacing is tighter than $\Delta \Lambda $ $=(\delta ^{2}S_{\mathrm{tot}%
}/\delta \Lambda ^{2})^{-1/2}$, we can approximate the discrete spectrum by
a continuous one. In Planck units, $\Delta \Lambda \sim \Lambda H$, and so
for the observed universe, where $H^{2}\sim \Lambda \sim 10^{-122}$, $\Delta
\Lambda \sim 10^{-186}$.

When we $S_{\mathrm{tot}}$ with respect to the matter and metric fields we
get the usual Einstein equations, but extremizing $S_{\mathrm{tot}}$ with
respect to $\Lambda $ gives a new field equation, $\delta S_{\mathrm{tot}%
}/\delta \Lambda =0$, equivalent to 
\begin{equation}
\frac{\,\mathrm{d}S_{\mathrm{cl}}(\Lambda )}{\,\mathrm{d}\Lambda }=0.
\label{eq:lambda}
\end{equation}%
where $S_{\mathrm{cl}}(\Lambda )$ is $S_{\mathrm{tot}}$ evaluated with the
matter and metric fields obeying their classical field equations. Eq. (\ref%
{eq:lambda}) is an additional field equation that links $\Lambda $ to the
other properties of the observable universe. Eq. (\ref{eq:lambda}) must hold
if the observable universe is to display an approximately classical
evolution, but there is no guarantee that a solution of the other Einstein
equations will also satisfy Eq. (\ref{eq:lambda}).

Since Eq. (\ref{eq:lambda}) restricts the possible classical universes, it
leads to testable predictions. When evaluated for our universe, we will find
that the observed value of $\Lambda $ determines the spatial curvature.
There is also now a simple argument for why $\Lambda \sim t_{\mathrm{U}%
}^{-2} $ is natural. Equation (\ref{eq:lambda}) is equivalent to 
\begin{equation}
\int_{\mathcal{M}}|g|^{\frac{1}{2}}\,\mathrm{d}^{4}x=\frac{1}{2}%
\int_{\partial \mathcal{M}}|\gamma |^{\frac{1}{2}}\left[ N^{\mu \nu }%
\mathcal{H}_{\mu \nu }+\Sigma _{a}\mathcal{P}^{a}\right] \,\mathrm{d}^{3}x.
\label{eq:phi}
\end{equation}%
The left-hand side is just the 4-volume, $V_{\mathcal{M}}$, of our spacetime 
$\mathcal{M}$. The right-hand side is a `holographic' term defined on the
boundary (of area $A_{\partial \mathcal{M}},$ say). Now, $N^{\mu \nu }%
\mathcal{H}_{\mu \nu }=N^{\mu \nu }\delta \gamma _{\mu \nu }/\delta \Lambda
\sim O(\Lambda ^{-1}\mathrm{tr}\,N)$ and $\Sigma _{a}\mathcal{P}^{a}$ is of
similar order of magnitude or smaller. Cosmologically, $\mathrm{tr}N\sim
O(H) $ where $H$ is the Hubble rate (with $H(t_{U})\equiv $ $H_{0}$ today),
and so the right-hand side of Eq.(\ref{eq:phi}) is $O(\Lambda
^{-1}H_{0}A_{\partial \mathcal{M}})$. So, we expect solutions of Eq.(\ref%
{eq:phi}) to have $\Lambda \sim O(H_{0})A_{\partial \mathcal{M}}/V_{\mathcal{%
M}}$. Typically, $H_{0}\sim A_{\partial \mathcal{M}}/V_{\mathcal{M}}$ and $%
H_{0}^{-1}$ is determined by $t_{\Lambda }=\Lambda ^{-1/2}$ and the age of
the universe $t_{\mathrm{U}}$. Eq.(\ref{eq:phi}) links the values of $%
t_{\Lambda }$ and $t_{\mathrm{U}}$ and, in the absence of fine-tunings, we
predict $t_{\Lambda }\sim O(t_{\mathrm{U}})$ and hence $\Lambda \sim
O(1)t_{U}^{-2}$ $\sim 10^{-122}$ in Planck units. If Eq.(\ref{eq:phi})
admits a classical solution, then the classical value of the effective $%
\Lambda $ will have the observed magnitude, $O(t_{U}^{-2})\sim 10^{-122},$
without any fine-tuning. The scale $t_{U}$ appears without fine tuning
because it defines \textit{our} past light cone which restricts the
variations in the action to be causal.

\bigskip We now apply our proposal to a Friedmann cosmology with metric: 
\begin{equation*}
\,\mathrm{d}s^{2}=a^{2}(\tau )\left[ -\,\mathrm{d}\tau
^{2}+(1+kx^{2}/4)^{-2}\,\mathrm{d}x^{i}\,\mathrm{d}x^{i}\right] ,
\end{equation*}%
where $k$ determines the spatial curvature. The matter is a perfect fluid
with pressure $P$, and density $\rho $.

The total action is 
\begin{equation*}
S_{\mathrm{tot}}=S_{\mathrm{EH}}+S_{\mathrm{\Lambda }}+S_{\mathrm{GHY}%
}^{(u)}+S_{\mathrm{m}}+\dots ,
\end{equation*}%
where $S_{\mathrm{EH}}$ is the Einstein-Hilbert action, $S_{\Lambda
}=-\kappa ^{-1}\int_{\mathcal{M}}\,\mathrm{d}^{4}x\sqrt{-g}\Lambda $, $S_{%
\mathrm{GHY}}^{(u)}$ is the Gibbons-Hawking-York boundary term on the
past-light cone boundary, $\partial \mathcal{M}_{u}$, of $\mathcal{M}$, and $%
S_{\mathrm{m}}$ is the matter action; the dots (...) represent boundary
terms on the initial hypersurface $\partial \mathcal{M}_{I}$. We assume that
all data on $\partial \mathcal{M}_{I}$ is fixed with respect to $\Lambda $
and so the variation of any $\partial \mathcal{M}_{I}$ surface terms with
respect to $\Lambda $ vanishes identically \cite{note1}.

We evaluate $S_{\mathrm{tot}}$ with matter and metric fields obeying their
classical field equations and find $S_{\mathrm{tot}}=S_{\mathrm{cl}}$ is 
\begin{equation*}
S_{\mathrm{cl}}=\frac{4\pi }{3}\int_{0}^{\tau _{0}}a^{4}(\tau )(\tau
_{0}-\tau )^{3}\left[ \kappa ^{-1}\Gamma -P_{\mathrm{eff}}(a)\right] \,%
\mathrm{d}\tau .
\end{equation*}%
The observer is at $\tau =\tau _{0}$, the total pressure is $P_{\mathrm{eff}%
}=P_{\mathrm{m}}-\mathcal{L}_{\mathrm{m}}$ and

\begin{equation*}
\Gamma \equiv (k/a^{2})[2/3+\tau /(\tau _{0}-\tau )].
\end{equation*}%
The dominant contributions to $P_{\mathrm{eff}}$ come from baryons, $P_{%
\mathrm{eff}}\approx -\mathcal{L}_{\mathrm{baryons}}=\zeta _{\mathrm{b}}\rho
_{\mathrm{baryons}}$, where $\zeta _{\mathrm{b}}\approx 1/2\ $in the
chiral-bag model for baryon structure \cite{longpaper}.

$S_{\mathrm{cl}}$ has no explicit $\Lambda $ dependence: all dependence on $%
\Lambda $ is encoded in the scale factor $a(\tau )$. We define $\delta \ln
a/\delta \lambda =\mathcal{A}(\tau )$ where $\mathcal{A}(\tau )$ is found by
perturbing Einstein's equations with respect to $\Lambda $ and requiring $%
\delta \ln a/\delta \Lambda =0$ initially to obtain 
\begin{equation*}
\mathcal{A}(\tau )=\frac{a(\tau )H(\tau )}{6}\int_{0}^{\tau }\frac{\,\mathrm{%
d}\tau ^{\ast }}{H^{2}(\tau ^{\ast })},
\end{equation*}%
where $H=a_{,\tau }/a^{2}$ is the Hubble parameter. We can now calculate $\,%
\mathrm{d}S_{\mathrm{cl}}/\,\mathrm{d}\Lambda =0$ and find that 
\begin{equation}
k=\frac{\kappa \int_{0}^{\tau _{0}}(\tau _{0}-\tau )^{3}a^{4}\zeta _{\mathrm{%
b}}\rho _{\mathrm{b}}\mathcal{A}(\tau )\,\mathrm{d}\tau }{\int_{0}^{\tau
_{0}}a^{2}(\tau )(\tau _{0}-\tau )^{2}(4(\tau _{0}-\tau )+6\tau )\mathcal{A}%
(\tau )\,\mathrm{d}\tau }.  \label{eq:const}
\end{equation}%
The right-hand side is positive and so only universes with $k>0$ can obey
Eq. (\ref{eq:lambda}) and possess a classical limit in a quantum cosmology.

Here, $k$ is the \textit{average} spatial curvature within the past light
cone, so $k>0$ applies to the observable universe -- not the whole-space
time -- and does \emph{not} require a closed global topology. For fixed $k$
and $\tau _{0}$ (and fixed initial conditions for the matter), Eq. (\ref%
{eq:const}) is an implicit equation for $\Lambda $ and predicts a relation
between $\Lambda $, $k$ and $\tau _{0}.$ If we measure $\Lambda $ and $\tau
_{0}$ we can predict the spatial curvature $k=k(\Lambda )$. For fixed $\tau
_{0}$, increasing $\Lambda $ requires smaller $k$ and for fixed $k$, $%
\Lambda $ decreases as $\tau _{0}$ increases.

Astronomers have measured our observation time, $\tau _{0}$, and the value
of $\Lambda ,$ but only have bounds on $k$. It is usual to express $k$ by
the dimensionless parameter $\Omega _{\mathrm{k0}}\equiv -k/a_{0}H_{0}^{2}$.
For our universe, taking $\Omega _{\Lambda 0}=0.73$, and a baryon density $%
\Omega _{\mathrm{b0}}=0.0423$, and CMB temperature $T_{\mathrm{CMB}}=2.725\,%
\mathrm{K}$, with $\zeta _{\mathrm{b}}=0.5$, we predict: 
\begin{equation*}
\Omega _{\mathrm{k0}}=-0.0055.
\end{equation*}%
This is consistent with the current 95\% CI of $\Omega _{\mathrm{k0}}\in
(-0.0133,+0.0084)$ \cite{Komatsu:2010fb}. Soon, data taken by the Planck CMB
satellite, together with constraints from baryon acoustic oscillations and $%
H_{0},$will test this precise prediction of $\Omega _{\mathrm{k0}}$.

There are other wider consequences of our scenario. At any given location
and time, the wave function of the universe is dominated by a classical
history in which $\Lambda $ takes a single constant value. Hence, no
classical time-evolution of $\Lambda $ can be observed. Yet the history that
dominates, and its associated $\Lambda $ value, changes at different
observation times. We see a history with $\Lambda =\Lambda _{1}$, but an
observer in our past would see a different history with $\Lambda =$ $\Lambda
_{2}>\Lambda _{1}$. For measurements of $\Lambda _{1}$ and $\Lambda _{2}$ to
be compared, information would have to be sent from one history to another.
At the level of classical physics this cannot be done. Observers will see a
history consistent with the constant $\Lambda $ given by Eq.(\ref{eq:phi})%
\cite{hist}. Crucially, this includes registering all previous measurements
of $\Lambda $ as being consistent with $\Lambda =\Lambda _{1}$. Therefore,
we do not see the past as an observer in the past would see it \cite{note2}.

Our simple extension of Einstein's theory therefore has striking
consequences: it explains the observed value of $\Lambda $, predicts the
curvature parameter of the universe, and paints a new picture of quantum
cosmological history.


\begin{thebibliography}{99}
\bibitem{units} We use Planck units in which we set $G=c=\hbar =1.$All times
are therefore given in units of the Planck time: $t_{pl}=(G\hbar
/c^{5})^{1/2}=5.4\times 10^{-44}s$.
\bibitem{note1} This implies that the form of Eq. (\ref{eq:lambda}) is
independent of this choice of the $\partial \mathcal{M}_{I}$ terms.
\bibitem{hist} Classically, this movement from one history to another has no
directly detectable consequences. From a quantum perspective, the wave
function is dominated by a superposition of histories with a small spread in 
$\Lambda $ of $\Delta \Lambda =(\delta ^{2}S_{\mathrm{tot}}/\delta \Lambda
^{2})^{-1/2}$. This superposition could give rise to new effects if a system
were sensitive to shifts of $O(\Delta \Lambda )$. However, with $\Omega
_{\Lambda 0}\sim O(1)$, $\Delta \Lambda /\Lambda \sim \Lambda ^{1/2}\sim
10^{-60}\ll 1$ but this effect looks undetectably small.
\bibitem{note2} This behaviour arises as a direct consequence of having
taken $\mathcal{M}$ to be the observer's causal past which, in turn, was
necessary to preserve causality when $\Lambda $ was promoted from a
parameter to a field.
\end{thebibliography}
\end{document}